\documentclass[lettersize,journal,10pt,]{IEEEtran}

\usepackage{xtab,booktabs}
\usepackage{import}
\usepackage[T1]{fontenc}
\usepackage{amsmath,amsthm}
\usepackage[normalem]{ulem}
\usepackage{algorithmic,algorithm}
\usepackage{multirow}
\usepackage{pgfplots}
\usepackage{threeparttable}
\usepackage{array}
\usepackage{graphicx,color}
\usepackage{listings}
\usepackage[labelfont=bf,font=small,skip=3pt]{caption}
\usepackage{pifont}
\usepackage{url}
\usepackage{array}
\usepackage{amsfonts}
\usepackage[cal=boondoxo]{mathalfa}
\usepackage{float}

\usepackage{framed}
\usepackage{soul}
\usepackage{xcolor}
\usepackage{tikz}
\usepackage{pgf}
\usepackage{fancyvrb}
\let\labelindent\relax
\usepackage{enumitem}
\usepackage{mathtools,amsthm}
\usepackage{cuted}
\usepackage{stackengine}
\usepackage{booktabs}
\usepackage{diagbox}
\usetikzlibrary{positioning}
\usetikzlibrary{decorations.pathreplacing}
\usetikzlibrary{3d} 
\usepackage{xspace}
\usepackage[utf8]{inputenc}
\usepackage{sidecap}
\usepackage{subcaption}
\usepackage{tikz}
\usepackage{fancybox}
\usepackage{eucal}
\usetikzlibrary{decorations.pathreplacing}
\usepackage{balance}
\usepackage{longtable,lscape}
\usepackage{threeparttablex}
\usepackage{tabularray}
\usepackage{hyperref}
\usetikzlibrary{snakes}
\usetikzlibrary{calc}
\usetikzlibrary{automata, positioning, arrows}
\usetikzlibrary{shapes.geometric}
\usetikzlibrary{arrows.meta}

\definecolor{bln_blue}{HTML}{00A8FF}
\definecolor{bln_red}{HTML}{c23616}
\definecolor{bln_green}{HTML}{16A085}
\definecolor{bln_magenta}{HTML}{9B59B6}
\newsavebox\mybox
\newsavebox\xbox
\newsavebox\ybox

\DeclareMathAlphabet\mathbfcal{OMS}{cmsy}{b}{n}

\renewcommand{\sout}[1]{}

\usepackage{xstring}
\newcommand{\mypara}[1]{
\vspace{.3em}
\noindent{\bf \IfEndWith{#1}{.}{#1}{\IfEndWith{#1}{?}{#1}{#1.}}}
}

\AtBeginDocument{}
\newcounter{question}[section]
\newcounter{mydefinition}[section]
\newcommand{\ie}{\textit{i}.\textit{e}.}
\newcommand{\eg}{\textit{e}.\textit{g}.}

\definecolor{mygray}{gray}{0.2}

\definecolor{tablegray}{gray}{0.9}
\lstdefinelanguage
    [x32]{Assembler}     
    [x86masm]{Assembler} 
    {morekeywords={movl,addl,cmpl,CMPXCHG16B,JRCXZ,LODSQ,MOVSXD, %
                  POPFQ,PUSHFQ,SCASQ,STOSQ,IRETQ,RDTSCP,SWAPGS,CALLQ,LEAVEQ,RETQ, %
                  rax,rdx,rcx,rbx,rsi,rdi,rsp,rbp, %
                  r8,r8d,r8w,r8b,r9,r9d,r9w,r9b, %
                  r10,r10d,r10w,r10b,r11,r11d,r11w,r11b, %
                  r12,r12d,r12w,r12b,r13,r13d,r13w,r13b, %
                  r14,r14d,r14w,r14b,r15,r15d,r15w,r15b}} 

\renewcommand{\tablename}{\footnotesize Table}

\def\Snospace~{\S{}}

\usepackage{relsize}

\newcommand{\cc}[1]{\mbox{\smaller[0.5]\texttt{#1}}}

\newcommand{\kenali}{\mbox{\textsc{Kenali}}\xspace}

\definecolor{dkgreen}{rgb}{0,0.6,0}
\definecolor{gray}{rgb}{0.5,0.5,0.5}
\definecolor{mauve}{rgb}{0.58,0,0.82}
\definecolor{mygray}{gray}{0.9}
\colorlet{lightblue}{blue!70}
\colorlet{lightred}{red!70}

\newcommand{\squishlist}{
\begin{itemize}[noitemsep,nolistsep]
  \setlength{\itemsep}{-0pt}
}
\newcommand{\squishend}{
  \end{itemize}
}

\usepackage{url}

\theoremstyle{definition}
\newtheorem{definition}{Definition}

\theoremstyle{definition}

\theoremstyle{definition}

\theoremstyle{definition}

\theoremstyle{definition}

\definecolor{bittersweet}{rgb}{1.0, 0.44, 0.37}
\definecolor{bleudefrance}{rgb}{0.19, 0.55, 0.91}

\begin{document}
\author{
    \IEEEauthorblockN{Zhilong Wang\IEEEauthorrefmark{1}, Haizhou Wang\IEEEauthorrefmark{1}, Hong Hu\IEEEauthorrefmark{1}, Peng Liu\IEEEauthorrefmark{1}}\\
    \IEEEauthorblockA{\IEEEauthorrefmark{1}Pennsylvania State University
    \\\{zzw169, hjw5074, honghu, pxl20\}@psu.edu}
}

\title{Identifying Non-Control Security-Critical Data through Program Dependence Learning}

\maketitle

\begin{abstract}

As control-flow protection gets widely deployed,
it is difficult for attackers to corrupt control-data and 
achieve control-flow hijacking. 
Instead, data-oriented attacks, which manipulate non-control data,
have been demonstrated to be feasible and powerful. 
In data-oriented attacks, a fundamental step is to identify 
non-control, security-critical data.
However, critical data identification processes are not scalable in previous works, because they mainly rely on tedious human efforts to identify critical data.
To address this issue, we propose a novel approach that 
combines traditional program analysis with deep learning.
At a higher level, by examining how analysts identify critical data, we first propose 
dynamic analysis algorithms to identify the program semantics (and features) that are correlated with the impact of a critical data. 
Then, motivated by the unique challenges in the critical data identification task, we formalize the distinguishing features 
and use customized program dependence graphs (PDG) to embed the features.    
Different from previous works using deep learning to learn basic program semantics, this paper adopts a special neural network architecture that can capture the long dependency paths 
(in the PDG), through which a critical variable propagates its impact.  
We have implemented a fully-automatic toolchain and conducted comprehensive evaluations. 
According to the evaluations, our model can achieve 90\% accuracy. 
The toolchain uncovers 80 potential critical variables in Google FuzzBench. In addition, we demonstrate the harmfulness of the exploits using the identified critical variables by simulating 7 data-oriented attacks through GDB.   

\end{abstract}


\section{Introduction}
\label{sec:intro}
\footnote{A previous version of this paper was posted on arXiv with number 2108.12071}
As control-flow protection mechanisms get 
mature~\cite{original-cfi,pi-cfi,bin-cfi,patharmor,ucfi} and widely
deployed~\cite{ms-cfi,clang-cfi,android-cfi,chromium-cfi}, it becomes
difficult for attackers to corrupt control data, like return addresses
or function pointers, to launch control-flow hijacking
attacks~\cite{rop,jop,coop,ret2libc}.
Attackers are prompted to search for remaining, novel hacking vectors,
and they have recently found potentials from data-oriented
attacks~\cite{data-attack,dop,bop,pewny2019steroids,cheng2019exploitation,ye2023viper,ahmed2023not}.
In a data-oriented attack, attackers leverage memory corruption vulnerabilities (\eg, buffer
overflow and use-after-free) to modify some \textit{non-control, 
security-critical} data to affect the program execution. 
The corrupted data enables the attacker to attain escalated privileges, 
authentication bypass, and/or more permissions, without being thwarted by control-flow protection mechanisms.
For example, \autoref{code:php} shows a code snippet from program \cc{php} which contains 
a critical variable \cc{disabled\_functions} disclosed in this work.  
Loaded from a configuration file~\textendash~\cc{php.ini}, this variable carries disabled \cc{php} functions selected by the administrator, among which could include security-sensitive functions.
Consequently, an attacker can execute these disabled functions by modifying the value of \cc{disabled\_functions}. 
For another example, \autoref{code:bit} shows a code snippet related to critical data \cc{aclp} in program \cc{proftpd}. A malicious user can bypass the authentication check by modifying \cc{aclp}.
%
%
In this paper, we will use the term \textit{critical data}
and the term \textit{critical variables} interchangeably. 


\begin{figure}[t] 
  \begin{lstlisting}[language=C,
      emph={zend_disable_functions,zend_ini_string_ex,zend_disable_function,zend_hash_str_del},
      caption={\textbf{An example of newly uncovered critical data~in~\texttt{php}.} \texttt{disable\_functions} are list of function names read from configuration file~\textendash~php.ini, which allows user to disable certain functions(\eg, \texttt{exec}, \texttt{shell\_exec}) to be invoked by php scripts.},label=code:php,captionpos=b]
    int main() { 
      char *disabled_functions =zend_ini_string_ex(...);
      zend_disable_functions(disabled_functions);
    }
    //zend_disable_functions -> ... -> zend_hash_str_del
    zend_result zend_hash_str_del(HashTable *ht, char *str, size_t len){ 
      while (idx != HT_INVALID_IDX) {
        p = HT_HASH_TO_BUCKET(ht, idx);
        if ( zend_string_equals_cstr(p->key,str,len)){
          _zend_hash_del_el_ex(ht, idx, p, prev);
        }
        idx = Z_NEXT(p->val);
      } 
    }
  \end{lstlisting}
\end{figure}
\begin{figure}[t]
\begin{lstlisting}[language=C,
    emph={setup_env,login_check_limits,check_limit_deny,check_user_access},
    caption={\textbf{An example of newly uncovered critical~data~in~\texttt{proftpd}.} \cc{aclp} determines whether the program accepts a user authentication.},label=code:bit,captionpos=b]
    int setup_env( ... ) { ...
        aclp = login_check_limits(conf, 0, 1, &i); ...
        if (c == NULL && aclp == 0) { 
          goto auth_failure; } 
    }
    int login_check_limits( ... ){ ...
        res = (res || rres); return res;
    }
  \end{lstlisting}
\end{figure}


In response to the increasing interests in data-oriented attacks,
researchers have developed various defense mechanisms. 
Since full memory-safety that protects all program 
data~\cite{dfi,softbound,cets,cyclone,ccured, safe-c} introduces
unacceptable runtime overhead, recent proposals mainly focus on
protecting merely critical data~\cite{xmp,dynpta,kdfi}. 

Nevertheless, the prerequisite of these defenses 
is whether one can proactively identify critical data. 
Unfortunately, existing works mainly rely on human 
to {\em manually} identify critical data, whose scalability is limited when screening 
large code bases or porting to new applications.
To confirm a critical variable $v$, one need to 
understand the program semantics of not only the code statements containing $v$, but 
also the code statements that have control/data dependency 
relationships. 
There are literature about data-oriented attack~\cite{data-attack} and its defense~\cite{xmp, schlesinger2014modular, dynpta}, but research on ``automatic identification of critical data'' can be rarely found.

\begin{figure}
  \centering
\includegraphics[width=1\columnwidth]{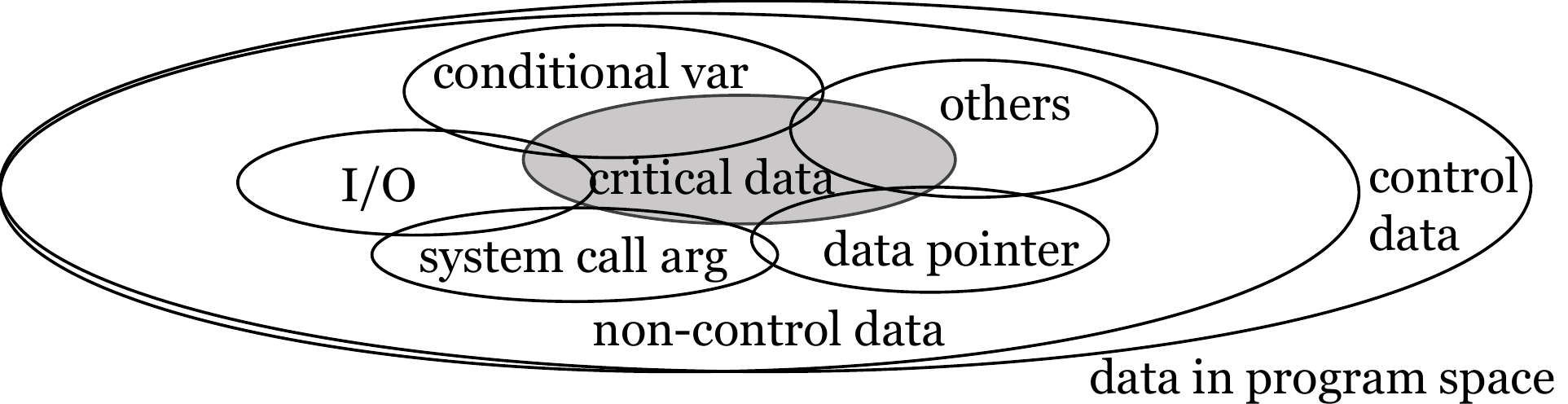}
  \caption{\footnotesize \textbf{Different categories of data in program space}.}
  \label{fig:noncontrol}
\end{figure}

This work aims to address this gap and 
propose an automatic workflow to identify 
critical variable candidates in various programs. 
A fundamental challenge of this research problem is that critical data are 
highly specific to each program and show no {\em explicit generic} patterns.  
Firstly, critical data could be any of the widely known 
data categories (shown in \autoref{fig:noncontrol}).
Secondly, criteria used by analysts to confirm critical data 
rely on {\em high-level program semantics}, which cannot be captured by existing 
lower-level program analysis tools (\eg, symbolic execution, taint analysis, etc.).

In order to address this fundamental challenge, we started with examining  
how analysts identify critical data. In particular, 
we observe that real-world security analysts are using
certain {\em distinguishing features} to identify critical data.  
As we will shortly illustrate in \autoref{fig:defaultroot} and \autoref{code:res} in \autoref{sec:charact}, 
all the distinguishing features are associated with 
the program behaviors (\eg,  ``critical functions are forbidden'' in \autoref{code:php}, ``user is blocked'' in \autoref{code:bit}) 
{\em dominated} by a critical variable. 
Inspired by this {\bf insight}, 
to establish a general workflow for identifying critical variables, we 
formalize such {\bf program behaviors dominated by a variable} 
through {\bf program state transition sequences}
predefined by programmers. 

Based on this key observation, we propose a 3-phase workflow: 
an $\mathcal{E}xecutor$, a $\mathcal{D}iff$ algorithm, and a $\mathcal{M}easure$ algorithm. 
The $\mathcal{E}xecutor$ produces different program state transition sequences by mutating each interested variable; the $\mathcal{D}iff$ algorithm analyzes the sequences and collects the program states dominated by the variable; and the $\mathcal{M}easure$ algorithm will confirm the criticalness of the variable by learning and measuring the states. 
To be more specific and detailed, the workflow starts with a program and a set of inputs:
Firstly, the $\mathcal{E}xecutor$ executes the program with an input that can trigger a variable $v$ in the program to produce a state transition sequence, where the initial value $x_0$ of $v$ is recorded. 
In the following executions with the same input, we mutate the 
value of $v$ to other values $\{x_i|1 \leq i \leq k\}$ just before the variable is used.
Note that the $\mathcal{E}xecutor$ customizes a fuzzing framework to mutate $v$, and monitors 
how it changes the state transition sequence.  
Secondly, the $\mathcal{D}iff$ algorithm analyzes and compares 
the sequences of program states to identify 
the variables that are data dependent on $v$, the basic blocks that are control 
dependent on $v$, and some other useful features. 
Thirdly, the $\mathcal{M}easure$ algorithm will appropriately 
measure the features identified during the second phase. 


Ideally, critical variables could be identified were the features appropriately
measured, but if the ways to measure the features are specific to each 
victim program, the 3-phase workflow could suffer from poor 
{\em generalization ability} (i.e., the ability to handle 
previously-unseen programs).
Therefore,
we employ deep learning (DL) to learn the decision boundaries between 
the program state transition sequences (as well as their differences) 
attributed to a critical variable 
and those attributed to a non-critical variable.  
Previous works have provided compelling evidence that thoughtfully designed 
neural networks are able to learn high-level program semantics~\cite{feng2020codebert,guo2020graphcodebert,allamanis2017learning,bubeck2023sparks}.   

However, in case of the critical variable identification task, we found that 
the existing learning strategies (e.g., pretrained models (CodeBert~\cite{feng2020codebert},  GraphCodeBert~\cite{guo2020graphcodebert}), GGNN~\cite{allamanis2017learning}, RNN~\cite{chua2017neural,guo2019deepvsa,shin2015recognizing,LQY21} and etc.) did not 
work very well: they are not devised to capture the   
{\bf long-range} control/data dependencies characterizing 
the impacts of a critical variable. 
To cross this barrier, our learning strategy uses dependence trees to hold 
all the extracted features, especially the long-range control/data dependencies. 
In addition, we choose a neural architecture (i.e. Tree-LSTM) that 
can naturally ``remember information for long periods of time'' and learn 
implicit patterns from long-range dependencies. 

We integrated the trained DL model into our fully-automatic toolchain. 
We used our tool to answer two questions. 
\ding{182} Can it re-discover the critical variables identified in previous works 
and discover unknown critical variables in these previous investigated programs? 
We found that our tool is able to rediscover 8 out of the 9 critical variables 
previously-investigated in \cite{data-attack} and \cite{flowstitch}, 
and find quite a few potentially critical variables (see \autoref{tab:rediscovery}). 
\ding{183} Can it uncover critical variables in other types of programs?   
Our tool conducted analysis on the FuzzBench~\cite{metzman2021fuzzbench}, which 
holds a variety of different types of programs.  
Our tool uncovers 80 potentially critical variables. 
In summary, our main contributions are as follows. 

\begin{itemize}[leftmargin=1.7em]

\item We developed an automatic toolchain which 
employs a new, formalized, deep-learning-assisted method to assist analysts in  
identifying critical variables.  
To our best knowledge, it is the first toolchain towards automatic identification of critical data in real world programs. 
\item We implemented and thoroughly evaluated our toolchain. 
Our deep learning model achieves 90\% accuracy. 
Our tool uncovers 80 potentially critical variables 
in Google FuzzBench. In addition, we have launched proof-of-concept simulated attacks on a subset of the 80 variables through GDB.
Against the previously-investigated programs, our tool is able to rediscover the 8 out of 9 previously-discovered critical variables. 
\item This work also creates the first publicly available, labeled dataset in the research area of critical variable discovery.   
\end{itemize} 

\section{Background}
\label{sec:back}

\subsection{Control data and non-control data}
\noindent\textbf{Control data} ``specifies the target location of a branch instruction. By changing control data, an attacker can arbitrarily change the control flow of an application. Examples of control data are return addresses and function pointers.
In contrast, \textbf{Non-control data} never contains the target address for a control transfer. In certain cases, however, it may influence the control flow of an application. For instance, a conditional branch may depend on the value of non-control data"~\cite{vogl2014dynamic}.

\subsection{Control flow hijacking attacks vs. data-only attacks}
\noindent\textbf{Control-flow hijacking attacks} exploit memory corruption vulnerabilities to divert program execution away from the intended control flow~\cite{burow2017control}.
A typical control hijacking attack starts by corrupting a pointer (\ie, control data) to an attacker-supplied malicious data, which we refer to as the payload~\cite{otgonbaatar2015evaluating}.
As the percentage of control data in program is small~\cite{kuznetzov2018code}, mature control-flow protection mechanisms~\cite{original-cfi,pi-cfi,bin-cfi,patharmor,ucfi} can effectively prevent control flow hijacking attacks with acceptable overhead and are widely deployed~\cite{ms-cfi,clang-cfi,android-cfi,chromium-cfi}. 

\noindent\textbf{Data-only attacks} 
exploit computer programs without altering any control data.
Instead, they manipulate certain security-critical non-control data (``critical data'' for short). Data-only attack can achieve many malicious goals, such as arbitrary code execution~\cite{safemode,emet-bypass,cfg-bypass}, authentication bypass and information leakage~\cite{heartbleed,data-attack}. 
Essentially, attackers exploit vulnerabilities to 
manipulate critical data to achieve their malicious purposes.
For example, by corrupting an authorization flag in the stack frame, 
an attacker can log into an SSH server as the root user
without providing any correct credentials~\cite{data-attack}. 
As most of data in program's are non-control data, it is not practical to protect all of them within reasonable overhead.
Despite the large amount of non-control data, few of them are interested by attackers. Attack surface can be greatly reduced if the small amount of ``critical data'' can be protected.



\section{Problem and Overview}
\label{sec:insights}
\begin{lrbox}{\mybox}
    \begin{minipage}{0.90\columnwidth}
\begin{lstlisting}[language=C, 
emph={DefaultRoot, User, Groyo,PathDenyFilter, DenyFilter, UseSendfile,Group,MaxInstances,ControlsEngine,ControlsACLs,ControlsSocketACL},
emphstyle={\color{bittersweet}},
morekeywords={LogFormat,UseIPv6,ControlsLog, SystemLog},
keywordstyle={\color{bleudefrance}},
stringstyle={},
xleftmargin=.00\columnwidth, 
    xrightmargin=.00\columnwidth,
% caption={Configuration variables from ProFTPD.},
label=code:config,captionpos=b]
    /* configuration variables of ProFTPD. 
        load from ProFTPD.conf */
    DefaultRoot         ~/Downloads
    UseSendfile         off
    ControlsACLs        all allow user root
    DenyFilter          777
    PathDenyFilter      \.jpg$  
    UseIPv6             off       
    LogFormat           "%h %l %u %t \"%r\" %s %b"
    SystemLog           /var/log/ProFTPD.log
\end{lstlisting}
  \end{minipage}
  \end{lrbox}

  \begin{lrbox}{\xbox}
\begin{minipage}{1.10\columnwidth}
\begin{lstlisting}[language=C,
    emph={path, c,DefaultRoot,default_root, cwd, new_cwd, full_path},
    emphstyle={\color{bittersweet}},
    keywordstyle={},
    stringstyle={},
    xleftmargin=-.02\columnwidth, 
    xrightmargin=.02\columnwidth,
    % caption={Program logics that dependent on DefaultRoot.},
    label=code:rootdep,captionpos=b]
    static const char *get_default_root(pool *p) {
      c = find_config( main_server->conf,  CONF_PARAM,  
              "DefaultRoot", FALSE);
      path = c->argv[0];
      if (strncmp(path, "/", 2) == 0) { ... }
    }
    static int setup_env(pool *p, char *user) {
      default_root = get_default_root(session.pool);
      if (pr_auth_chroot(default_root) < 0) { ... }
      if (strncmp(session.cwd, default_root, strlen(default_root)) == 0) { 
        new_cwd = &session.cwd[strlen(default_root)];
        sstrncpy(&session.cwd[1],new_cwd,sizeof(session.cwd));
      }
      dir_check_full(p, cmd, G_NONE, session.cwd,NULL);
    }
    xaset_t *get_dir_ctxt(pool *p, char *dir_path) { 
      full_path = pdircat(p, session.chroot_path, session.cwd, dir_path, NULL);
    }
\end{lstlisting}
  \end{minipage}
    \end{lrbox}
    \begin{lrbox}{\ybox}
      \begin{minipage}{0.90\columnwidth}
\begin{lstlisting}[language=C,
    emph={SystemLog,c, path, fmt},
    emphstyle={\color{bleudefrance}},
    keywordstyle={},
    stringstyle={},
    xleftmargin=.00\columnwidth, 
    xrightmargin=.00\columnwidth,
    % caption={Program logics that dependent on SystemLog.},
    label=code:logdep,captionpos=b]
    static void log_postparse_ev(
      const void *event_data, void *user_data) {
      c = find_config(main_server->conf, CONF_PARAM, "SystemLog", FALSE);
      path = c->argv[0];
      res = log_opensyslog(path);
    }
\end{lstlisting}
  \end{minipage}
  \end{lrbox}

\subsection{Critical data}
\label{sec:whatis}
Data-only attacks (defined by~\cite{schlesinger2014modular}) leverage memory corruption bugs to directly or indirectly manipulate a ``critical data'' to control the program behaviors.
An indirect manipulation can be achieved through a data pointer, array index, or gadget chain. 
\begin{center}
\footnotesize
{\tt bug(s)} $\xRightarrow{\text{step~1}}$ {\tt critical-data} $\xRightarrow{\text{step~2}}${\tt malicious-behavior}
\end{center}
The manipulated ``critical data'' then enables the attacker to attain escalated privilege, authentication bypass, or more permissions. 
We observed that although step~1 usually breaks the integrity of program data flow, step~2 does not.
Considering that no clear definition of ``critical data'' has been proposed yet.
Readers can easily lump the ``critical data''  with ``sensitive data'' (such as user input and system call arguments~\cite{ahmed2023not}) and ``vulnerable data'' (such as data pointers, and array index). To avoid vagueness, we propose the following definition:

\tikzset{
node distance=3.5cm, 
every state/.style={thick}, 
initial text=$\ \ $, 
global scale/.style={
scale=#1,
every node/.append style={scale=#1}
},
directedge/.style = {->},
ass/.style = {-{>[width=2pt 2]}},
imp/.style = {-{Latex[open]}},
proh/.style = {{Bar[sep=1pt]}-{Bar[sep=1pt]}}
}

\vspace{0.5em}
\noindent\textbf{Preliminary.} A program state ($s_i$) is the set of values of variables at a given point during a program execution. 
It is a snapshot of the current state of computation~\cite{waite1987c,ball1996efficient}. 
A state transition $\langle s_i, s_j \rangle$ corresponds to the execution of a code statement 
and a state transition sequence represents the changes of the program state along with its execution. We use $\left(s_n\right)_{n \in \mathbb{N}}$ to denote a sequence, where $\mathbb{N}$ is a set of natural numbers.

\vspace{0.5em}
\begin{definition}[\bf Security-critical Data]
\label{def:var}
Let $\left(s_n\right)_{n \in \mathbb{N}}$ denote the program state transition sequence (predefined by programmers) of a particular program under a specific input.
Suppose that a variable $v$ with a value of $x_1$ is used in $s_{i}$. 
Then, changing the value of $v$ to $x_2$ at $s_{i}$ will lead to another predefined program state transition sequence $\left(s_n^2\right)_{n \in \mathbb{M}}$ in resulting execution. 
Namely, $v$ dominates the two state transition sequences.

 \begin{tikzpicture}[global scale=0.5, ]
    \node[state, scale=0.8, initial, initial where=left] (1) {$ $};
    \node[state, scale=0.8, right of=1] (2) {$ $};
    \node[state, scale=0.8, right of=2] (3) {$ $};
    \node[state, scale=0.8, right of=3] (4) {$ $};
    \node[state, scale=0.8, right of=4] (5) {$ $};
    \node[state, scale=0.8, below right of=2] (6) {$ $};
    \node[state, scale=0.8, below right of=3] (7) {$ $};
    \node[state, scale=0.8, below right of=4] (8) {$ $};
    \draw
        (1) node[above=0.2, scale=1.3] {\large $s_{i-1}$}
        (2) node[above=0.2,scale=1.3] {\large $s_{i}$}
        (3) node[above=0.2,scale=1.3] {\large $s_{i+1}$}
        (4) node[above=0.2,scale=1.3] {\large $s_{i+2}$}
        (5) node[above=0.2,scale=1.3] {\large $s_{i+3}$}
        (6) node[above=0.2,scale=1.3] {\large $s_{i+1}^2$}
        (7) node[above=0.2,scale=1.3] {\large $s_{i+2}^2$}
        (8) node[above=0.2,scale=1.3] {\large $s_{i+3}^2$}
        (5) node[right=0.3, scale=1.1] {\Large sequence $\left(s_n\right)_{n \in \mathbb{N}}$}
        (8) node[right=0.3, scale=1.1] {\Large sequence $\left(s_n^2\right)_{n \in \mathbb{M}}$}
        (2) node[below=0.8, scale=1.5] (9) {\Large $v$}
        (1) edge[directedge, above] node[scale=0.8,text width=1.2cm]{} (2)   
        (2) edge[directedge, above] node[scale=0.8,text width=1.2cm]{} (3)  
        (3) edge[directedge, above] node[scale=0.8,text width=1.6cm]{} (4)
        (4) edge[directedge, above right = 0.1] node[scale=0.8,text width=1cm]{} (5) 
        (2) edge[directedge, above] node[scale=0.8]{} (6)
        (6) edge[directedge, below] node[scale=0.8]{} (7)  
        (7) edge[directedge, below] node[scale=0.8]{} (8)
        (9) edge[dash dot, ->, below] node[scale=0.8]{} (2)
        ;
    \end{tikzpicture}  

\noindent Compared with $\forall s_j \in \left(s_n\right)_{n \in \mathbb{N}}$, if $\exists s_{k}^2 \in \left(s_n^2\right)_{n \in \mathbb{M}}$ has the property of enabling the attacker to attain escalated privilege, authentication bypass, or more permissions, then we say $s_{k}^2$ is a security-critical state and $v$ is a \textit{security-critical data variable}. Otherwise $v$ is a \textit{security-noncritical data variable}. We denote this special property (\textsc{E}scalated privilege, \textsc{A}uthentication bypass, or more \textsc{P}ermissions) as the \textbf{\textsc{EAP} property}. 
\end{definition}

\textsc{Caveat 1.} Note that the definition does not consider the existence of memory corruption vulnerabilities, which is used in step~1 of data-only attacks. That is because critical variable discovery is orthogonal to memory corruption bug finding.
Variables that trigger memory corruption are usually not considered as security-critical data because the memory corruption results in a program state undefined by programmers. Neither is a corrupted pointer that is leveraged in complex data-only attacks to further corrupt a security-critical data to reach security-critical state, because the corrupted pointer dominates the security-critical state through a state transition path that is not predefined by programmers. 
A manipulated variable, which can only trigger program crush or DoS, is not considered  as security-critical data either unless the crush/DoS is predefined program state. 

\textsc{Caveat 2.} Since the \textsc{EAP} property achieved by a critical variable is through a predefined state transition path, the defense mechanisms that are based on control flow integrity and data flow integrity will not affect the variable's criticalness.

Based on Definition~\ref{def:var}, we can confirm that a variable is security-critical in the following scenario: if the manipulation of the value of the variable is a sufficient condition \textbf{by itself} to enable a data-only attack (\ie, reaching security-critical state), it is a critical variable.


\subsection{Distinguishing features of critical data}
\label{sec:charact}

The related literature~\cite {data-attack,flowstitch,dop} shows that real-world security analysts usually use certain semantic-level features to assess whether a program state transition sequence may satisfy the EAP property. 
\autoref{fig:defaultroot} and \autoref{code:res} show two pairs of critical and non-critical variables, respectively.
Config file in \texttt{ProFTPD} has tens of configuration variables that are loaded into memory and control FTP server's behaviors. Box-1 in \autoref{fig:defaultroot} lists a few of them. 
The red color variables are identified as critical variable. For example, modifiying \texttt{\color{bittersweet} UseSendfile}, \texttt{\color{bittersweet} PathDenyFilter} enable the attacker to access more files on the server.  
The blue variables are not because no security-critical program state is reached by modifiying them.

We take \texttt{\color{bittersweet} DefaultRoot} and \texttt{\color{bleudefrance} SystemLog} as examples.
Specifically, \texttt{DefaultRoot} restricts users to only certain directories. An attacker can have access to more files by simply manipulating it. 
Whereas the \texttt{SystemLog} specifies the location of the log files and modifying it does not give more permission to the user.
Since measuring program states that are not predefined is difficult (\eg,~overwrite some files by manipulating \texttt{\color{bleudefrance} SystemLog} here), we only consider the variable's impact (\ie, \textsc{EAP} properties) on the predefined program state transition paths.

\begin{figure*}
  \begin{tikzpicture}[
    node distance=2cm,
    startstop/.style={rectangle, rounded corners, minimum width=3cm, minimum height=1cm,text centered, draw=black, fill=red!30},
    process/.style={rectangle, minimum width=3cm, minimum height=1cm, text centered}, 
    io/.style={trapezium, trapezium left angle=70, trapezium right angle=110, minimum width=3cm, minimum height=1cm, text centered, draw=black, fill=blue!30},
    decision/.style={diamond, minimum width=3cm, minimum height=1cm, text centered, draw=black, fill=green!30},
    ]
  
    \node (node0) [process]                             {\usebox\mybox};
    \draw[draw=black,line width=0.25mm] ($(node0) + (-3.98, 0.68)$) rectangle ++(8, 0.26);
    \draw[draw=black,line width=0.25mm] ($(node0) + (-3.98, -1.32)$) rectangle ++(8, 0.26);
    \node (node1) [process, right of=node0, xshift=7.5cm, yshift=-1.4cm]            {\usebox\xbox};
    \node (node2) [process, below of=node0, xshift=0cm, yshift=-1.2cm]            {\usebox\ybox};
    \draw
      (node0) node[above=1.1, xshift=3.3cm, scale=1.0] {\small \tt Box-1}
      (node2) node[above=0.8, xshift=3.3cm, scale=1.0] {\small \tt Box-2}
      (node1) node[above=2.4, xshift=4.0cm, scale=1.0] {\small \tt Box-3}
      ($(node0) + (4, 0.7)$) edge[directedge, line width=0.25mm, dashed, above] node[scale=0.8,text width=1.6cm]{} ($(node0) + (6.0, 0.7)$)
      ($(node0) + (-1.5, -1.30)$) edge[directedge,line width=0.25mm, dashed, above] node[scale=0.8,text width=1.6cm]{} ($(node0) + (-1.5, -2.6)$)
      ;
  \end{tikzpicture}
  \caption{{\bf Code pieces dependent on two variable \texttt{DefaultRoot} and \texttt{SystemLog} from \texttt{ProFTPD} config files.} \texttt{DefaultRoot} restricts users to only certain directories, \texttt{SystemLog} specifies the path to output the logs.}
  \label{fig:defaultroot}
  \end{figure*}

Box-2 and Box-3 in \autoref{fig:defaultroot} show the code statements that are data dependent on \texttt{SystemLog} and \texttt{DefaultRoot} respectively. We colored the variables that are affected by the two variables.
In Box-2, \texttt{SystemLog} is loaded in \texttt{path} and then opened through the \texttt{log\_opensyslog()}. In Box-3, the \texttt{DefaultRoot} path firstly goes through 3 rounds of checks:
line-4 checks whether it is a root path; line-8 checks folder's access permission; line-13 checks compliance for other user defined constrains.
Secondly, the value from \texttt{DefaultRoot} path is used when responding to several FTP requests, \eg, CWD, PWD, and etc. \texttt{get\_dir\_context()} calculate the full path of the accessed file or folder. 


\autoref{code:res} shows two loop control variables. 
\texttt{\color{bittersweet} authenticated} is clearly critical. 
The value of \texttt{authenticated} comes from \texttt{auth \_password()}, which is dependent on username and password from the \texttt{packet\_read()}. 
\texttt{\color{bleudefrance} res} is a non-critical variable.
The value of \texttt{res} is the return value of \texttt{write()}~\textendash~a library in \texttt{Glibc}. \texttt{write()} return the number of bytes written if it successfully writes some data, otherwise return -1 when I/O error happens. 
Changing the value of \texttt{res} does not have any security impact.

The examples show that comparing with non-critical variables, a critical variable usually has bigger impact on the program, which can be reflected in different ways: changing execution path, changing the range of being-accessed resources and so on. 
Secondly, the critical variable could control the program through data dependence (\autoref{fig:defaultroot}) or control dependence (\autoref{code:res}).  
Thirdly, the value of critical data can be affected by the input directly or indirectly.
Besides these distinguishing features, we have observed some other characteristics of critical data in our dataset: 
(1) The distribution of critical variables is uneven across various program types and functions. Without context, it can be challenging to determine whether a variable in a general-purpose function or library is critical or not. 
For example it is difficult to determine whether a variable in library \texttt{jsoncpp} is critical, without knowing what is being parsed.
In contrast, program with specific business logic may contain easy-to-confirm critical data.
For example, network applications such as FTP servers usually have more critical variables. This observation is aligned with previous works~\textendash~most previous studies focus on network applications. 
(2) Critical data could fall into many data categories (shown in \autoref{fig:noncontrol}): I/O, data pointer, variables used in conditional statements, arguments of system calls, and etc. In each category, only a small portion is critical data. 
(3) Critical variables are most commonly involved in 
logical operations; they are rarely involved in 
complex arithmetic operations (\texttt{MUL}, \texttt{DIV}).


\subsection{{\bf The Research Problem}: How to check whether a variable is critical data?}
\label{sec:problem:researchproblem}


Security researchers have discussed or investigated the feasibility of using some of the aforementioned distinguishing features to narrow down the search space to find critical data~\cite{data-attack,ahmed2023not,ye2023viper}. For example, one could leverage the existing taint analysis techniques and a rule-based filter to identify variables that can be tainted by inputs, variables that propagate their taint to conditional statements (\ie, \texttt{CMP}, \texttt{TEST} etc.) or to arguments of system calls.
However, as mentioned by~\cite{data-attack}, the achieved accuracy is in general fairly low because many tainted variables are false positives.

{\bf Our Idea:} According to Definition~\ref{def:var}, a sufficient condition is the satisfaction of the \textsc{EAP} property in the program state tranistion sequences. 
Therefore, we can propose a more formalized workflow
to determine whether a variable is critical or not. Suppose a program has $k$-state transition sequences (corresponding to $k$ executions) {\em after} the value of the variable is changed. Let \{$v_i| 1 \leq i \leq k\}$ denote the $k$ distinct values. Let's denote the $k$ different state transition sequences $\{\left(s_n^i\right)_{n \in \mathbb{N}}|1 \leq i \leq k\}$. 

Let $\mathcal{D}{iff}(\left(s_n^1\right)_{n \in \mathbb{N}},\left(s_n^2\right)_{n \in \mathbb{N}}, \cdots \left(s_n^k\right)_{n \in \mathbb{N}})$ denote an algorithm to analyze the $k$-state transition sequences and return a set of differences ($\mathbb{D}$) among the sequences. $s_d \in \mathbb{D}$ denotes a state that appeared in at least one sequence and not in one another sequence. 
We use $\mathcal{M}{easure}(\mathbb{D})$ to denote the algorithm to measure the likelihood that the differences $\mathbb{D}$ will cause \textsc{EAP} related security issues and then determine whether the variable is security-critical.

The criticalness of the variable is reflected by the differences between the state transition sequences.
In some extreme cases, no matter what value was assigned to a variable, never is the program state transition path changed, we can confidently say that this variable is not critical.
In some other cases, if the state transition sequence is changed accordingly, we can use $\mathcal{D}{iff}$ algorithm to collect the differences between the sequences, and then use $\mathcal{M}{easure}$ algorithm to measure the differences.
Critical variables can be identified if the differences are appropriately measured.
Especially, the measurements should focus on the \textsc{EAP} property in the set of different states.
In the following sections, we will focus on how to design and implement the two algorithms, \ie, $\mathcal{D}{iff}$ and $\mathcal{M}easure$.

\subsection{Challenges and our solutions}
\label{sec:challenge}
Firstly, we need to generate state transition sequences for executions.
Program state transition sequence can be generated through symbolic execution or concrete execution. Currently, the symbolic execution faces several challenges (\eg, path explosion, failed constraint solving and etc).
Analyzing concrete executions is very efficient and accurate, however may suffer from coverage issues. We decided to collect the program states through concrete executions, which are more practical in real-world programs. 

Secondly, based on the high-level design discussed in \autoref{sec:problem:researchproblem}, for each variable $v$ we want to confirm (to be critical or not), we need to launch several executions with $v$ of different values. 
To achieve this goal, we directly alter the variable to desired values during different executions.

Thirdly, the program states and its transition include the context information of the program's control flow, data flow, data values etc. We need to select useful information from state transition sequences as the input of $\mathcal{D}{iff}$ algorithm. 
Altering a variable could result in different execution paths and different values of taint-propagated variables.
Therefore, $\mathcal{D}{iff}$ needs to be able to analyze both the affected control flow and variables. We adopt a customized program dependence graph to represent how a variable affects the control flow and data flow. 

Fourthly, one challenge we faced when developing $\mathcal{M}easure$ algorithm is that \texttt{EAP} property is related to high-level program semantics and cannot be learned and analyzed by traditional program analysis.
For example, whether the \texttt{authenticated} in \autoref{code:res} is critical is related to the context of the piece of code. Only after the analyst understands that the loop is used for authentication can he/she confirm the critical variable. 
However, even if an analyst had a good understanding of such authentication flags, it is still challenging to devise an explicit generic pattern that can decide whether a loop is related to authentication.
Some analysts could argue that he/she can infer the purpose of \texttt{authenticated} simply through its name. Actually, such ``infer-from-name'' is a commonly used strategy by human analysts to infer the functionality of a variable or piece of code.  
Although this trick works well for some cases, it cannot be used to identify critical data with unconventional names or form general heuristic rules.

We adopt deep learning to learn program semantics related to \texttt{EAP} property. 
That is, we take a DL approach to implement the $\mathcal{M}easure$ algorithm which trains a DNN model to recognize semantics patterns unique to critical variables~\textendash~such patterns are automatically learned from the aforementioned distinguishing features. 
Previous works have provided compelling evidence that thoughtfully designed neural networks are able to learn some of the high-level semantics information from a program~\cite{feng2020codebert,guo2020graphcodebert,allamanis2017learning,bubeck2023sparks}.
However, in solving our problem, we found that the intuitive neural network architecture (GNN-based model) failed to learn enough information due to two serious issues: over-smoothing (\ie, the model cannot distinguish data examples with different labels when it stacks many layers to learn an implicit pattern  
embedded in a {\bf long} path in a PDG) and node feature diluting. 
Our experiments (\autoref{sec:baseline}) demonstrate that several other widely adopted models in program analysis are also incapable.
In order to resolve these issues, we convert each data example from a graph structure to a feature-preserving tree structure. We call such trees a dependence tree. 
In addition, we find that a Tree-LSTM ~\cite{tai2015improved}, 
which is a generic version of a regular LSTM model, can effectively learn
from long dependence paths.

\begin{figure}[]
\begin{lstlisting}[
      emph={authenticated, log_file},
      emphstyle={\color{bittersweet}},
      stringstyle={},
      keywordstyle={},
      % caption={A authentication flag in OpenSSH.},
      label=code:auth,captionpos=b]
    int authenticated = 0;
    while (!authenticated) {
      type = packet_read();
      if (auth_password(user, password)){
        authenticated =1;  }
      if (authenticated) break;
    }
\end{lstlisting}
\begin{lstlisting}[language=C,
    emph={res},
    emphstyle={\color{bleudefrance}},
    stringstyle={},
    keywordstyle={},
    captionpos=b]
    res = write(forensic_logfd,fm->fm_msg,fm->fm_msglen);
    while (res < 0) {
      pr_signals_handle();
      res=write(forensic_logfd,fm->fm_msg,fm->fm_msglen);
    }
\end{lstlisting}
\caption{{\bf Two loop control variables.} \texttt{authenticated} in \texttt{OpenSSH} is an authentication flag; \texttt{res} in \texttt{ProFTPD} denotes whether a file writting success.}
\label{code:res}
\end{figure}

Finally, besides the \texttt{EAP} property, there are some other characteristics which we can use to confirm a critical variable.  
For example, as mentioned earlier a critical variable usually has a bigger impact on the program execution when comparing with non-critical ones. Therefore, the size of $\mathbb{D}$, \eg, counting the number of affected variables and number of the affected basic blocks (BBs), is also an useful feature. 
For another example, as shown in \autoref{code:res}, the source where a variable comes from also includes useful information.
With all the features, we leverage a DL model to learn and draw the decision boundary between critical and non-critical variables.

\subsection{Approach overview}
\label{sec:overview}

\begin{figure*}[t]
    \centering
    \includegraphics[width=2.05\columnwidth]{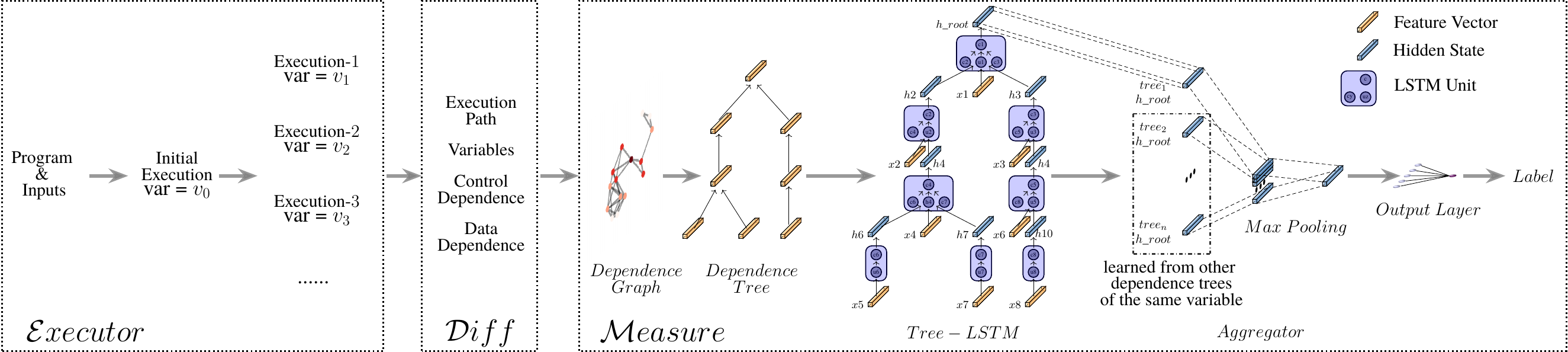}
    \caption{\footnotesize \textbf{Approach overview.}}
    \label{fig:overview}
  \end{figure*}

\autoref{fig:overview} shows the detailed workflow, which consists of 3 phases: an $\mathcal{E}xecutor$, a $\mathcal{D}iff$ algorithm, and a $\mathcal{M}easure$ algorithm.
The workflow starts with a program (either source code or binary) and a set of inputs ( generated through Fuzz tools~\cite{afl,yun2018qsym}).
Firstly, the $\mathcal{E}xecutor$ executes the program several times with an input, which can trigger a list of variables in the program.
For each variable ($var$) in the list, we monitor the value ($v_0$) of $var$ during the initial execution. In the following executions, we mutate the value of $var$ to other values $\{v_i|1 \leq i \leq k\}$ just before the variable is used.
Secondly, $\mathcal{D}iff$ algorithm analyzes and compares the set of executions to identify the variables ($\mathbb{D}_v$) that are data dependent on $var$, BBs ($\mathbb{D}_b$) that are control dependent on $var$, together with some other useful features.

Thirdly, the $\mathcal{M}easure$ will first embed $\mathbb{D}_v$ and $\mathbb{D}_b$ into a program dependence graph (PDG). The data/control dependence relationships are obtained by dynamically analyzing the executions. 
Besides, through feature engineering, which is presented in \autoref{sec:data}, we obtain i) a particular set of features for each node in the PDG and ii) the type information for each edge.
Fourthly, for each node in the PDG, our tool-chain first vectorizes 
each individual node feature, and then concatenates all the individual vectors into a single {\em concatenated} vector. 
 Fifth, our tool-chain converts the PDG 
to a feature-preserving dependence tree (see details in \autoref{sec:lstm} and \autoref{fig:dfg}). By ``feature-preserving", we mean that 
if node A in the graph is mapped to node B in the tree, then node B will 
have the same concatenated vector as node A. 

Sixth, the unrolled Tree-LSTM neural network, shown at the center of \autoref{fig:overview}, 
is trained to learn the implicit pattern 
embedded in the dependence tree. 
Intuitively, our model leverages the capability of LSTM models  
in learning long range temporal dependencies as follows: 
The dependence tree often holds a set of long paths; each path 
starts at a leave node and ends at the root node; such long paths 
do hold long range temporal dependencies. 
Nevertheless, due to the tree-structure, the long paths 
should not be treated 
as linear chains. For example, since the first path 
and the second path shown in \autoref{fig:overview} 
have a common sub-sequence, namely $x_4$ $x_2$ $x_1$, 
the LSTM unit that processes $x_4$ should incorporate not only 
the information propagated along the first path (\ie, $x_5$) 
but also the second (\ie, $x_7$). 
This is why we adopt a Tree-LSTM model. 
The Tree-LSTM unit which processes $x_4$ will 
``fuse'' the information propagated along both paths.  

Seventh, the last hidden state outputted by the Tree-LSTM neural   
network is essentially the learned {\bf representation} for 
the input dependence tree. 
If the corresponding variable $x$ has other live-variables, we will 
also learn a separate representation from each of them. 
Eighth, 
we use a pooling layer to aggregate all the learned representations 
for variable $x$.  
Finally, the output layer will predict a label for $x$
using the aggregated representation.

\section{Workflow Design}
\label{sec:data}
The goal of $\mathcal{E}xecutor$ and $\mathcal{D}iff$ is to analyze the program and identify the program states that can be affected by a target variable. 
If we can precisely obtain the set of BBs and variables dependent on the variable, we can confirm which part of the program states are affected by the variables.
In our design, we choose dynamic analysis instead of static analysis to obtain such dependence relationships because static analysis is not very scalable to analyze an inter-procedural dependence relationship.
\subsection{Uncover the control dependence}\label{sec:cdp}
A single dynamic execution is not enough to uncover all possible states that are related to the target variable. 
Therefore, we choose to execute the programs multiple times, change the value of the target variable, and then monitor how the changed variable affects the execution. We design an algorithm to uncover the BBs that are dominated by a variable $V$ in program $\mathbf{P}$. 
The formalized the algorithm will be provided in the supplemental material.
Specifically, the algorithm tries to find BBs that are control-dependent on the $V$ by mutating the value of $V$ during the execution. After repeating this process, we will obtain a set of execution paths.
Those BBs that appeared in one execution path but not in another are confirmed to be control-dependent on $V$. 
The quality of mutated values from $\mathcal{V}$ can affect the quality of our algorithm. A ``bad'' value set could fail to uncover important paths. 
In our tool, we directly adopt the popular mutation strategies used in fuzzing test~\cite{zhu2022fuzzing}: 1) flip each bit in a variable, respectively; 2) flip its value based on conditional statements that use the variable. For example, if a variable with the value of \texttt{0x5} in \texttt{\%eax} is used in statements \texttt{cmp \$0x8,\%eax; jne 401180;}, we change its value to \texttt{0x8} so that the conditional branch can be flipped.

\subsection{Uncover the data dependence} \label{sec:under}


We analyze the execution traces to uncover all other variables that are data-dependent on the variable $V$. 
There are two kinds of data dependencies to consider. 
1) \textit{Direct dependencies} exist when a variable is assigned to, or used to calculate the value of, another variable. 
2) \textit{Implicit dependencies} are one kind of data dependencies that are difficult to analyze even with dynamic analysis, which is widely discussed in taint analysis~\cite{newsome2005dynamic, kang2011dta}. 
One type of implicit dependency exists when a variable affects an array index or a pointer. For example, in the following code, 
\cc{c} is used to address the source operand, then affects the value of \cc{a} which will finally control the critical behavior. 
\begin{center}
  \small
  \texttt{a = b[c]; if (a == 10)\{~critical behavior~}\}
\end{center}

Another type of implicit dependencies  
exist when a variable affect the control flow which might in turn affects other variables~\cite{newsome2005dynamic, kang2011dta}. 
In \autoref{code:php}, \cc{disabled\_functions} contains the names of the functions to be disabled. 
Following the data dependency from \cc{disabled\_functions}, the dependency path finally reaches the return value of \cc{zend\_string\_equals\_cstr()}. 
However, the \cc{\_zend\_hash} \cc{\_del\_el\_ex()} deletes a function 
based on \cc{idx} which control-dependent on 
the return value of \cc{zend\_string\_equals\_cstr}. 
As a result, we cannot connect variable \cc{disabled\_functions} with 
its ``critical impact'' through a path in a DDG or a CDG.

To capture the first type of implicit dependencies, we can add edges to connect the array index or pointer with the accessed variable. 
To capture the second type, we combine the data and control dependencies in an unified graph. 
However, nodes in traditional DDGs and CDGs are usually defined as an instruction or an statement, which are not 
variable-specific. 
Therefore, we customized their definitions as follows:
\begin{definition}[Customized DDG]
\label{def:cddg}
In a customized DDG $DDG(N,E)$, each node in $N$
represents variables used in 
predicate expressions (or instructions) of
the program, while each edge in $E$ represents a data dependency
between two nodes in $N$. 
One node $n_1$ is data-dependent on another $n_0$ iif. $n_0$ is used to calculate the value of $n_1$; 
or $n_0$ is directly copied to $n_1$; 
or $n_0$ is used to address $n_1$. 
\end{definition}
\begin{definition}[Customized CDG]
\label{def:ccdg}
In a customized CDG $CDG(N,E)$, each nodes in $N$
represents variables used in predicate expressions (or instructions) of
the program, while each edge in $E$ represents a control dependency
between two nodes in $N$. 
Providing that $n_0$ and $n_1$ are used in statement $s_0$ and $s_1$, respectively. $n_1$ is control-dependent on $n_0$ if
and only if statement $s_1$ is control-dependent on $s_0$ and $n_1$ is assigned with new values in $s_1$ and $n_0$ is used in the conditional statement of $s_0$.
\end{definition}

The two new definitions enable us to represent data and control dependencies in a unified PDG where nodes represent variables, 
and edges represent data/control dependencies between variables. 
The new definition enable us to connect the \cc{disabled\_functions} and its ``critical impact'' through the path: {\cc{disabled\_functions}}
{$\xrightarrow{}$\cc{zend\_string\_equals\_cstr}$\dashrightarrow$\cc{idx}$\to$\cc{\_zend\_hash\_del\_el\_ex}}, where $\to$ and $\dashrightarrow$ represent data and control dependency, respectively.

\subsection{Feature engineering}

The PDG only presents the data/control dependency relationship among variables, we need to select 
more features specific to critical variable identification. 

\mypara{Semantic-aware feature selection}
To enable the $\mathcal{M}easure$ to learn the semantics of a critical variable, 
we select two features: {\bf Opcode:} the opcode of instructions or expressions carry essential semantics. 
{\bf Type of dependency:} data and control dependencies 
are two different types carrying different semantics. 

\mypara{Feature embedding} 
As mentioned in \autoref{sec:overview}, 
before we convert each PDG to a dependence tree, 
the selected features are firstly {\em embedded} in 
each PDG graph either as node or edge features.  
We distinguish \textbf{two} types of nodes in a PDG graph: 
i) an \textit{$o$-node} represents an operation
performed on a variable, whose node features are the 
operation's opcode. 
ii) a \textit{$v$-node} represents a \textit{live}-variable
whose node features include its data/control 
dependency measurement (ie, $|\mathbb{D}_v|$ and $|\mathbb{D}_b|$). 

We define \textbf{four} types of edges to represent different kinds of dependencies. 
Specifically, a \textit{$d$-edge} describes an explicit data dependency;
an \textit{$i$-edge} represents an implicit dependency;
a \textit{$c$-edge} indicates an control dependency; 
and an \textit{$r$-edge} describes the redefinition relationship. 
The first three types of dependencies have been defined in 
Definition~\ref{def:ccdg} and Definition~\ref{def:cddg}.
An $r$-edge indicates whether two or more live-variables belong to the 
same variable in code. 

\begin{figure}[t]
  \centering
  \includegraphics[width=0.90\columnwidth]{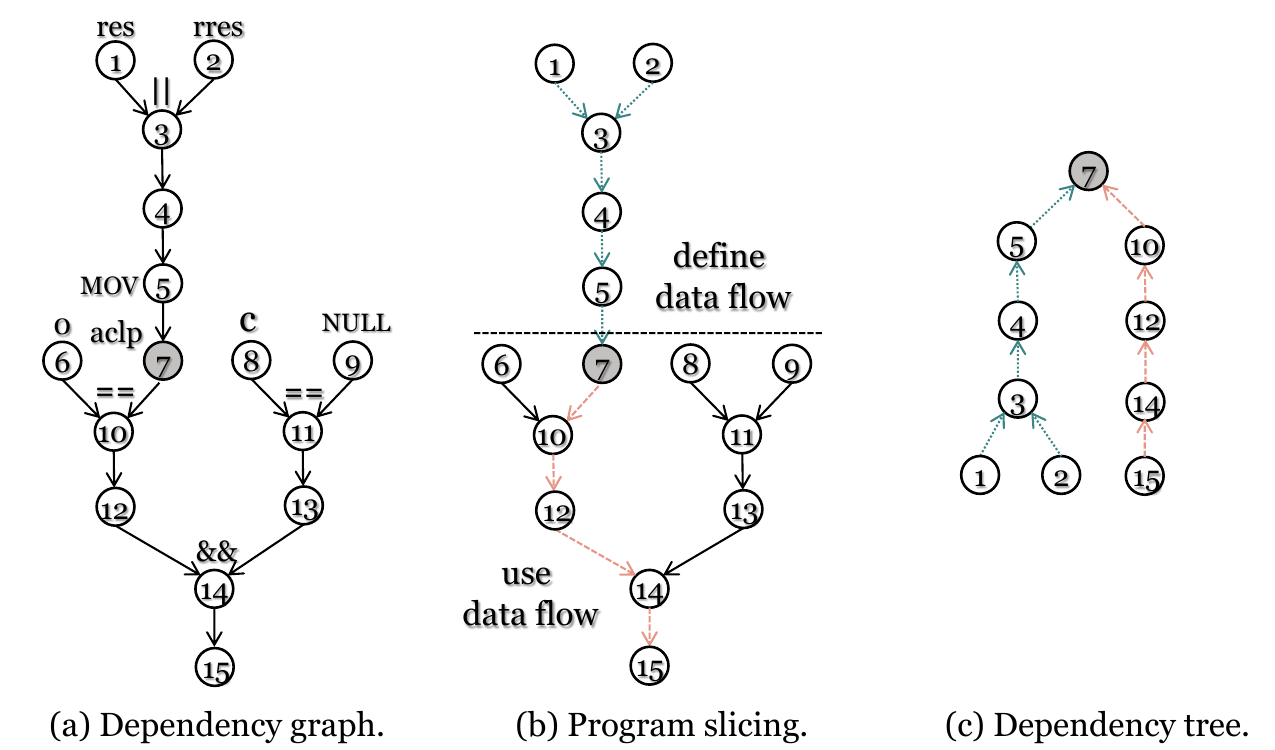}
  \caption{Dependence graphs and tree built from the execution trace of \autoref{code:bit} for variable \texttt{aclp}.}
  \label{fig:dfg}
\end{figure}

\autoref{fig:dfg}(a) shows a PDG generated 
from the code shown in \autoref{code:bit}. The 
graph is specific to variable \cc{aclp}.  
In the graph, nodes 1-5 and 7 are generated 
from the operations in Line 7.
Node 6 and Nodes 8-15 are generated from the operations in Line 3.
Among all these nodes, nodes 3, 5, 10, 11, and 14 
are $o$-nodes, and the others are $v$-nodes. 

\subsection{Tool-chain and data sample labeling} 
This section briefly introduce the implementation of the tool chain to generate data, whose detail is included in the supplemental material.

\label{sec:datagen}
\mypara{An automatic PDG construction tool} We developed an automatic tool-chain that traces the execution of a program and builds the PDG for each variable that appeared in the execution trace. 
The tool-chain includes a Intel Pin tool~\cite{luk2005pin}, which collects various runtime information including executed instructions and accessed memory addresses; 
an $\mathcal{E}{executor}$, which is a modified version of AFL, will automatically mutate a target variable based on strategies in \autoref{sec:cdp}, and generate multiple execution traces; a $\mathcal{D}{iff}$ implementation which analyze and compare multiple traces, then generates PDGs (each PDG is corresponding to an define-use of the variable) for~each~interested~variable.

\noindent{\bf Data sample labeling.}  
Although various critical variables are 
reported in the literature~\cite{data-attack,flowstitch}, 
\textbf{no} labeled dataset is currently available. 
To get such a dataset, we firstly selected 10 previously-investigated programs (\ie, \cc{bftpd}, \cc{ghttpd}, \cc{telnet}, \cc{vsftpd},
\cc{proftpd}, \cc{nginx}, \cc{nullhttpd}, \cc{sshd}, \cc{sudo}, \cc{wu-ftpd})~\cite{data-attack,flowstitch,dop}.
Secondly, we asked two experts to read the source code of the 10 programs 
and manually labelled a set of variables. 
A variable can only be labeled as ``critical" if a data-only attack 
can be launched by modifying the variable. 
Thirdly, we used a customized compiler to propagate the label information from 
the source code to the binaries.  
Note that the source code is \textbf{only used for labeling}. 
Finally, we utilized our tool-chain to 
generate the PDGs for each labeled variable. 
As a result, we generated a training set containing 6,000 PDG samples (each PDG sample corresponds to one variable use instance), with 3,340 critical 
and 2,660 non-critical samples.
We tried our best to increase the diversity of the types of critical data, so that our dataset includes critical data of decision-making, user inputs, crypto key, privilege control, configuration, etc.

\section{Neural Architecture Design}
\label{sec:gnnlstm}

We choose tree-structured long short-term memory model (Tree-LSTM)~\cite{tai2015improved} as the backbone of our neural network architecture. 
However, compared to GNN, Tree-LSTM is less widely used in the literature. 
Therefore, we will first explain a GNN-based neural architecture, and then explain why it is not suitable for our task. 
\begin{figure}[t]
    \centering
    \includegraphics[width=.8\columnwidth]{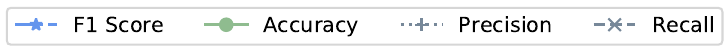}
    \begin{subfigure}[t]{0.49\columnwidth}
      \includegraphics[width=0.97\textwidth]{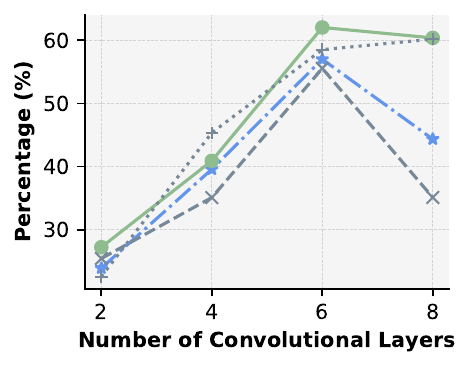}
      \caption{Results of RGCN}
      \label{fig:depth-gnn}
    \end{subfigure}
    \begin{subfigure}[t]{0.49\columnwidth}
      \includegraphics[width=\textwidth]{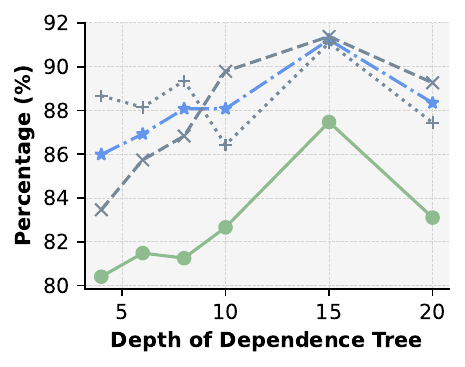}
      \caption{Results of Tree-LSTM}
      \label{fig:depth-lstm}
    \end{subfigure}
    \caption{\textbf{Comparison between GNN and Tree-LSTM}. We perform the evaluation with different number of layers and flow depths to find the optimal design choice.}
    \label{fig:depth}
\end{figure}

\subsection{Unsatisfying GNN} 
\label{sec:lstm_motivation}


Firstly, we designed a GNN(Graph Neural Network)-based neural architecture, which consists of three components:
1) a representation learner, which learns the representations for individual nodes in a PDG; 
2) an aggregator, which aggregates the individual node representations into a graph-level representation; 
3) a classifier, which predicts whether the learned graph representation is critical. 
\autoref{fig:depth-gnn} shows the F1 score and accuracy of the RGCN
model on identifying critical data. 
The x-axis indicates different configurations of the model, each with
a different number of hidden layers. 
We can see that the model performs better when we increase the number
of layers before reaching six, but the performance drops after six. 
The six-layer RGCN produces the best performance with 62.00\%
accuracy and 0.5698 F1 score. 

\mypara{Limitations of GNN}
\label{sec:lesson}
By analyzing the results in \autoref{fig:depth-gnn}, we find several
limitations of RGCN for our task.
On the one hand, RGCN models with fewer layers cannot effectively
learn features in long paths of a PDG. 
The number of layers in an RGCN determines the 
length of a (data/control) dependence path propagated
during node representation learning. 
The ``impact'' of a critical variable is often manifested in a fairly long 
dependence path, and such ``impact'' will be missed by 
``shallow'' RGCN models. 
On the other hand, simply increasing the number of layers is not helpful 
due to two reasons: the over-smoothing issue, and pervasive loops in the message-propagation path. 
First, GNN commonly suffers from the {\bf over-smoothing} issue where it cannot
distinguish data with different labels when the model stacks too many
layers (typically when $k>6$)~\cite{chen2020measuring,8578815}.  
Secondly, the many loops in the RGCN's message propagation path when learning the node features for 
PDG, which may {\bf dilute} the features of some nodes.

\subsection{Tree-LSTM for long range dependence learning}
\label{sec:lstm}


To address the limitations of GNN, we propose to use
Tree-LSTM~\cite{tai2015improved} to learn implicit patterns 
embedded in long dependency paths. 
Based on the characteristics of the dependencies important to us, 
we find that converting each PDG into a tree structure 
serves two purposes simultaneously: (1) each long path in the tree 
is inherently suited for LSTM learning which does not suffer from 
over-smoothing; (2) because the tree does not have any loops, and therefore the pervasive loops issue are no longer valid.
Based on this key insight, we adopt the Tree-LSTM neural architecture. 

{\em Extracting Dependence Trees from a PDG. }  
For each live variable, its dependencies fall into
two categories: 
i) {\bf Define-flow}, which is the data or control dependency
  involved in calculating a value assigned to the variable.
ii) {\bf Use-flow}, which is the data or control dependency involved in using the value of the variable.
Take \autoref{code:bit} as an example. 
Line~7 calculates the value of variable \cc{aclp}, which is a define-flow.
Line~3 uses variable \cc{aclp}, which is a use-flow.

With the definitions of define-flow and use-flow, we can extract 
dependence trees from a PDG as follows. 
Firstly, we identify the define-flow and use-flow for each
live-variable we are interested in.  
Starting from a particular live-variable $x$, 
we can generate two subgraphs by forwardly and backwardly slicing the PDG, respectively.
The blue path and the orange path in \autoref{fig:dfg}(b)
demonstrate the define-flow and the use-flow of live-variable
\cc{aclp}, respectively. 

Secondly, we flip the direction of the edges 
in the use-flow subgraph. As a result, the two subgraphs can be
merged into a tree with $x$ as the root node:  
\autoref{fig:dfg}(c) shows the resulting dependence tree. 
Next, we assign a different type to the flipped edges to 
distinguish define-flows and use-flows.
Thirdly, we transfer the node features and edge type information from the PDG to the dependence tree. 
Finally, we set up a threshold $k$ during the slicing to enforce a
reasonable depth for each dependence tree. 


{\em Adopting Tree-LSTM.}
As described in \autoref{fig:overview}, 
our Tree-LSTM design has three components: 
a tree representation learner, 
a representation aggregator,
and a classifier.  
The major differences between our Tree-LSTM model 
and a common LSTM model are as follows: 
(a) An unrolled LSTM neural network is a sequence of 
LSTM units, but our neural network is a tree
of LSTM units. 
(b) When a node has two or more children, the corresponding Tree-LSTM unit will have additional gates to ``fuse'' information propagated
from all of its children. 
Similar to a common LSTM model, the last hidden state 
outputted by our Tree-LSTM model is essentially the representation learned for the input dependence tree. 
After the representation learner obtains the representations for each live-variable corresponding to the same variable in code, 
we adopt the pooling layer to aggregate the learned representations.  
Then, we use an output layer to predict the label of the variable 
based on the aggregation results. 


\noindent\textsc{Noting.}
Even though there are some GNNs (\eg, GraphLSTM~\cite{peng2017cross}, GGNN~\cite{li2015gated}) that potentially deal with long-range dependencies. 
Our investigation shows that both of them are not capable. 
For example, the GraphLSTM~\cite{peng2017cross} is adopted to process \textit{document graphs}. 
However, a PDG is much more complex than a document graph, and it cannot be firstly converted to two directed acyclic document graphs and then learned by a GraphLSTM model.  

\section{Evaluation}
\label{sec:eval}



We aim to answer five questions:
\ding{182} Does our method perform well on the test dataset?
\ding{183} Is our method better than the baselines?
\ding{184} Can our method rediscover the critical variables found in previous works and discover previously unknown critical variables?
\ding{185} Can our method uncover previously unknown critical variables in other types of programs?

\subsection{Dataset statistics and model training}

\noindent{\bf Labeled Set.} We use the manually labeled variables in \autoref{sec:datagen} to generate 6,000 dependence trees (each tree is a data samples; each tree corresponds to a PDG), which contains 3,340 positive sample and 2,660 negative samples. 
The dependence trees then were used to evaluate our model. 
Technically speaking, the size of our dataset is chosen according to the size of other similar datasets for graph classification tasks.
Specifically, the sizes of most datasets in other graph classification tasks usually range from hundreds to thousands~\cite{graphsoat}. 
For example, the three most commonly used datasets in the public graph classification benchmark~\cite{graphsoat} (\ie, PROTEINS~\cite{gallicchio2019fast}, MUTAG~\cite{gallicchio2019fast}, and NCI1~\cite{gallicchio2020ring}) contain 1113, 4110 and 188 data samples, respectively.


\noindent{\bf Previously-investigated Set. } 
We used this dataset to evaluate our method's ability to rediscover critical variables (manually) found in previous works and discover previously unknown critical variables. This dataset includes six programs that were also used in previous works~\cite{data-attack,flowstitch}. We prepared inputs that can trigger previously-investigated variables and generated dependence trees for each triggered variable. Note that although this set shares programs with the labeled set, {\em no} variable is in both sets. 



\noindent{\bf Set of Other Program Types.} 
Most previous works have focused on critical variables in 
network server programs (e.g., \texttt{nginx}) and user privilege management programs (e.g., \texttt{sudo}). 
Characteristics and distribution of critical data in other types of programs are yet to be investigated.  
To address this gap, we analyzed programs from Google FuzzBench~\cite{metzman2021fuzzbench}. Our analysis aims to 1) determine if other types of programs also contain critical variables, and identify the characteristics of critical variables there; 2) gain a better understanding of the generalization ability of our tool. 
\mypara{Model Evaluation}
Model performance is calculated through 10-fold cross-validation. 
In each of 10 rounds, we choose the data examples from 9 out of 10 programs 
as the training set, and the data examples from the remaining program 
as the test set. {\bf In this way, each trained model is always tested on a previously \textit{unseen} program.}
We perform each training for 50 epochs on a machine with Intel
Xeon CPU E5-2650 and 64GB memory. 
\subsection{Our method's performance on the test set}

The result of our model is shown in \autoref{tab:explain}.
Specifically, our model successfully identified 2979 out of 3,340
critical variable use instances, and misclassified only 220 ({\bf 8.27\%} false positive rate) out of 2,660 non-critical variable use instances as critical. 
As a comparison, the GNN model merely achieved 62.00\%
accuracy and a F1 score of 0.5698 on the same dataset.
Note that the design model is very stable during training. 
We did not observe any symptom of over-fitting or accuracy fluctuating during the training process.


\begin{table}[t]
  \centering
  \captionsetup{justification=centering}
  \footnotesize
  \setlength{\tabcolsep}{5pt}
  \caption{\textbf{Comparison between our model and baselines.}}
  \label{tab:explain}
  \begin{tabular}{p{2.4cm}<{}cccc} 
  \toprule
  \bf{Model} &  \bf{Accuracy} & \bf{Precision}  & \bf{Recall} & \bf{F1} \\
  \midrule
  
  LSTM & 0.7465 & 0.8484 & 0.6871 & 0.7593 \\
  GGNN & 0.6525 & 0.6987 & 0.3703 & 0.4818 \\
  RGCN & 0.6200  & 0.5849 & 0.5553 & 0.5698\\
  GraphCodeBERT & 0.6823  & 0.7058 & 0.8000 & 0.7499\\
  \midrule
  {\bf Tree-LSTM (Ours)} & {\bf 0.9032}   & {\bf 0.9312} & {\bf 0.8919} & {\bf 0.9111} \\
  \bottomrule
  \end{tabular}
\end{table}

{\bf Remark on the false positives.} We manually inspect the false positives 
and find that they can be grouped into 6 categories. 
(a) Most false positives are in the general purpose functions with complex logic, for example, \cc{ngx\_vslprintf()} in the \cc{nginx} could be used to process critical strings or non-critical strings.
(b) Critical-data-related data, that cannot directly be used to launch an attack, but will impact the operation of critical variables. \eg, type of the encryption algorithm, length of a key.
(c) Some variables in the log functions or time process functions, that have complex dataflow. For example, a variable controlling the log level. For another example, time could be used to either verify whether a token is expired in authorization (\eg, \texttt{openssh}) or to be used in non-critical operation (\eg, simply a timestamp). 
(d) Array or string pointers. 
(e) Return values of some library functions (\eg, file descriptor) that could be critical but hard to be used to launch~attacks.

{\bf Remark on false negatives. } 
We find that the false negatives can be grouped into 2 categories:
(a) some triggered critical variables do not show critical behaviors in its execution path; and 
(b) some triggered critical variables (\eg, \cc{vsftpd}) with special patterns that are not possessed by samples in the training set. 

\subsection{Comparison with baselines}
\label{sec:baseline}
\vspace{-0.5mm} 
Since no previous solution targets the critical-data
identification problem, we choose 4 previously used deep-learning methods in achieving other program analysis objectives as the baselines.
We train these models with features extracted from these same data samples that are used in our model.

The first category of the models in the benchmark include RNN and its variants, such as LSTM are two popular neural networks in previous binary analysis works~\cite{chua2017neural,guo2019deepvsa,shin2015recognizing,LQY21}. 
We firstly trace instructions that are reachable through
forward and backward (PDG) slicing within $15$-hops from a variable. 
Then, we train the sequence models mentioned above using the traced instructions.
An instruction sequence is usually longer than 200 bytes, which can cause
RNN-based models suffering from gradient explosion and gradient vanishing. Therefore, we adopt a hierarchical LSTM that similar to the design adopted by DEEPVSA
\cite{guo2019deepvsa}, which contains two sets of LSTM cells to learn embedding on instruction and sequence levels separately.
The results in \autoref{tab:explain} show that our model outperforms the two sequence models in both accuracy and F1 score. 
It indicates that the critical-data identification problem cannot be well handled by an RNN/LSTM based model trained using instruction sequences.

Next, we tried a variant of graph neural networks (a 4-layer vanilla 
GGNN~\cite{li2015gated}), which are trained using the extracted PDGs.
The GGNN adopts gated recurrent units in massage-propagation to maintain long-range dependencies.
\autoref{tab:explain} shows that GGNN cannot
effectively identify critical data.
Baseline 5 is RGCN, which is discussed in \autoref{sec:lstm_motivation} with the best layer number 6. 

We also considered methods at source code level. GraphCodeBERT~\cite{feng2020codebert,guo2020graphcodebert} is a pre-trained model based on Transformer which directly learns source code representations through self-supervised training tasks and a large-scale unlabeled corpora. We used GraphCodeBERT as the starting point and train the downstream classification task~\textendash~critical variable classification. Each data example is a piece of source code resulted from program slicing of a variable. Unfortunately, the resulting DL model suffers from low accuracy (\eg, 68.23\%). After analyzing the false positives and negatives, we have identified that the low accuracy of the model is primarily due to inadequate feature engineering. The deep learning model is only able to recognize explicit patterns such as variable and function names, which are insufficient for the task at hand. Therefore, further feature engineering is required to improve the accuracy of the model. Our observation aligns with the conclusion of other researchers~\cite{zhang2023features}.

%


\subsection{Rediscovering the critical variables identified in previous works}

Two previous works \cite{data-attack,flowstitch} had manually discovered several critical variables. 
We leverage such ground truth to evaluate our our method. 
Note that we skip the programs that cannot be compiled in modern Linux, \eg, \texttt{httpdx}, and skip the variables that does not align with our definition~\ref{def:var}.

\begin{table*}[t]
  \captionsetup{justification=centering}
\caption{\textbf{Newly uncovered critical variables that were confirmed by GDB.} }
\label{tab:gdbconfirm}
\centering
\footnotesize
\setlength{\tabcolsep}{4pt}
\begin{tabular}{m{2.8cm}<{}m{3.8cm}<{}m{2.5cm}<{}m{3.8cm}<{}m{3cm}<{}}
\toprule
{\bf Program} & {\bf Function} & {\bf Variable}  & {\bf The Potential Attack} & {\bf GDB Break Point}  \\
\midrule
php\_php-fuzz-parser  &  \texttt{zend\_disable\_functions}  &  function\_list  &  permission escalation & zend\_API.c:3263 \\
php\_php-fuzz-parser & \texttt{virtual\_cwd\_main\_cwd\_init} & cwd & changing work directory  & zend\_virtual\_cwd.c:188 \\
nginx  &  \texttt{ngx\_process\_options}  &  cycle-confix.data &  config manipulation & nginx.c:987 \\
curl\_curl\_fuzzer\_http  &  \texttt{seturl}  &  path &  URL manipulation  & urlapi.c:852 \\
sqlite3\_ossfuzz &  \texttt{appendAllPathElements}  &  pPath->zOut &  config manipulation  & sqlite3.c:42001 \\
sqlite3\_ossfuzz &  \texttt{sqlite3BtreeEnter}  &  p->sharable &  permission escalation  & sqlite3.c:66648 \\
sqlite3\_ossfuzz &  \texttt{sqlite3TableLock}  &  iDb &  permission escalation  & sqlite3.c:115275 \\
\bottomrule
\end{tabular}
\end{table*}
\begin{table}[t]
  \captionsetup{justification=centering}
  \caption{\textbf{Critical variable re-discovery.} }
  \label{tab:rediscovery}
\begin{threeparttable}
\footnotesize
\begin{tabular}{p{0.7cm}<{\centering}p{2.3cm}<{\centering}p{2.5cm}<{\centering}p{0.9cm}<{\centering}}
\toprule \toprule
{\bf Program} & {\bf Function} & {\bf Variable} & {\bf Score\tnote{1}}  \\ 
\cmidrule[1pt]{1-4}

\multirow{2}{0.6cm}{wu-ftpd} & \texttt{getdatasock} & pw->pw\_uid &  0.9567  \\ 
\cmidrule[\lightrulewidth](lr){2-4}\addlinespace[0ex]
& \multicolumn{3}{p{7.2cm}<{\centering}}{{\bf any others:} pass:use\_accessfile, pass:pw->pw\_uid, dataconn:usedefault, mapping\_chdir:dir, do\_daemon:Bypass\_PID\_Files, pam\_check\_pass:pam\_error} \\
\cmidrule[0.8pt]{1-4} 

\multirow{3}{1cm}{openssh} & \texttt{auth2\_record\_key}  & authenticated &  0.9368   \\ 
& \texttt{userauth\_pubkey}  & authenticated &  0.9296  \\ 
& \texttt{platform\_sys\_dir\_uid} & uid   & 0.9287   \\ 
\cmidrule[\lightrulewidth](lr){2-4}\addlinespace[0ex]
& \multicolumn{3}{p{7.2cm}<{\centering}}{{\bf any others:} sshauthopt\_merge:additional->valid\_before,
auth\_authorise\_keyopts:now,
ssh\_digest\_final:l,
input\_service\_request:authctxt->success,
input\_service\_request:acceptit,
sshkey\_perm\_ok:st.st\_uid,
channel\_permit\_all:pset->num\_permitted\_user,
ssh\_digest\_final:digest,
wait\_until\_can\_do\_something:now,
prepare\_auth\_info\_file:success,
private2\_decrypt:check2} \\
\cmidrule[0.8pt]{1-4} 

\multirow{2}{1cm}{telnet} & \texttt{start\_login} & loginprg   & 0.9008\\ 
\cmidrule[\lightrulewidth](lr){2-4}\addlinespace[0ex]
& \multicolumn{3}{p{7.2cm}<{\centering}}{{\bf any others:} addarg:avs->argc} \\
\cmidrule[0.8pt]{1-4} 

\multirow{2}{1cm}{sudo} & \texttt{get\_user\_info} & ud->uid   & 0.9008  \\ 
\cmidrule[\lightrulewidth](lr){2-4}\addlinespace[0ex]
& \multicolumn{3}{p{7.2cm}<{\centering}}{{\bf any others:} 
sudo\_secure\_path:sb, sudoers\_lookup:match, 
check\_user:ret, 
env\_should\_keep:keepit, 
runaslist\_matches:rc, 
hostlist\_matches\_int:matched, 
host\_matches:rc, 
cmnd\_matches:rc, 
userpw\_matches:uid, user\_matches:rc, sudoers\_lookup\_check:cmnd\_match, sudo\_make\_pwitem:uid, user\_in\_group:gid, sudo\_auth\_approval:status } \\
\cmidrule[0.8pt]{1-4} 

\multirow{3}{1cm}{nginx} & \texttt{PKCS8\_pkey\_get0} & p8->pkey  & 0.9515 \\ 
& \texttt{ngx\_add\_path} & p  & 0.8844   \\ 
\cmidrule[\lightrulewidth](lr){2-4}\addlinespace[0ex]
& \multicolumn{3}{p{7.2cm}<{\centering}}{{\bf any others:} ssl\_cipher\_apply\_rule:rule, check\_pem:len, getrn:nn} \\
\cmidrule[0.8pt]{1-4} 

\multirow{2}{1cm}{nullhttpd} & \texttt{config\_read} & config.server\_base\_dir & 0.9474  \\ 
\cmidrule[\lightrulewidth](lr){2-4}\addlinespace[0ex]
& \multicolumn{3}{p{7.2cm}<{\centering}}{{\bf any others:} config\_read:founddir } \\
  \bottomrule
  \bottomrule
  \end{tabular}
    \begin{tablenotes}[leftmargin=0cm]
      \scriptsize
\item[1] confidence score reported by the model.
\item[2] any other critical variable with higher score, each potentially critical variable is presented with pair of (function name : variable).
    \end{tablenotes}
    \end{threeparttable}
\end{table}

\autoref{tab:rediscovery} shows the rediscovery results. 
Our tool-chain rediscovered 8 out of 9 critical variables (0.90 model confidence score is our decision boundary).
Besides, for each rediscovered critical variable, we manually analyzed all the variables in the same program that got higher confidence scores. 
Consequently, we found quite a few potential critical variables, shown in \autoref{tab:rediscovery}.
Due to the page limit, we will not dive deeper here but will post the details online instead.

\subsection{Uncovering critical variables in the programs of other types}
\label{sec:wild}

\begin{table}[t]
  \captionsetup{justification=centering}
\caption{\textbf{Critical variable discovery in programs of other types.}} 
\label{tab:newdiscovery}
\centering
\footnotesize
\setlength{\tabcolsep}{4pt}
\begin{threeparttable}
\begin{tabular}{p{1.8cm}<{\centering}p{1.8cm}<{\centering}p{1.8cm}<{\centering}p{1.8cm}<{\centering}}
\toprule
{\bf Program} &  {\bf Triggered}~\tnote{1}  & {\bf Predicted}~\tnote{1} & {\bf Confirmed}~\tnote{1} \\
\midrule
curl   & 158  & 25  & 19 \\
php    & 381  & 35  & 20 \\
sqlite3  & 799  & 56  & 41 \\
total   & 1338	& 116 &	80 \\
\bottomrule
\end{tabular}
\begin{tablenotes}
  \scriptsize
  \item[1] 
  {\bf Triggered}: number of variables that trigger by an execution; 
  {\bf Predicted}: number of critical variables predicted by our tool; 
  {\bf Confirmed}: number of potentially critical variables through our manual verification. 
\end{tablenotes}
\end{threeparttable}
\end{table}

Beyond the programs investigated in the previous works, we use Google FuzzBench, which holds various types of programs. 
We have two goals in this experiment. Firstly, we want to know whether 
our trained model has the needed generalization ability against other types of programs.
Secondly, the diversity of programs in Google FuzzBench helps us investigate the 
distribution of critical data in different types of programs. 
As mentioned in~\ref{sec:charact}, there are very few critical variables in data-parsing-centric 
programs and general-purpose library functions. 
Therefore, to save space, we exclude all such programs in FuzzBench from our table.
As a result, only 3 programs are reserved in our analysis: \texttt{curl} (a command line tool), \texttt{php}, and \texttt{sqlite3}. 
\autoref{tab:newdiscovery} shows the analysis results. Overall, we found {\bf 80} manually-confirmed critical variables that were previously unknown. 

{\bf PoC simulated attacks using GDB against the identified critical variables.} 
In real-world cyber operations, the analyst review process has two main steps: 1) use domain expertise to identify critical variables; 
2) use proof-of-concept (PoC) or simulated attacks to exploit the variables identified in Step 1. Because we don't have enough human resources to exploit all the variables listed in \autoref{tab:newdiscovery} and \autoref{tab:rediscovery}, we only took Step 2 against 
a {\em subset} of the variables which are most interesting.
Before there is a memory corruption CVE that we can leverage, we will not be able to craft a data-oriented PoC.  Therefore, we use GDB to {\bf simulate} such attacks. 
In particular, we {\bf simulate} the effect of exploiting a memory corruption bug by 
using GDB to directly modify the value of the variable 
targeted by the memory corruption exploit. 
In total, we have successfully simulated 7 proof-of-concept exploits using GDB in 4 programs. 

Let's take a closer look at the 7 involved programs in~\autoref{tab:gdbconfirm}:   
{\bf php:} The details of \texttt{disabled\_functions} has been shown \autoref{sec:intro} based on \autoref{code:php}.
\sout{in allows one to disable certain functions, whose name are read from the configuration file (php.ini). By modifying this variable, attacker can remove the security-critical functions, such as \texttt{exec()}, \texttt{shell\_exec()}, \texttt{curl\_exec()}) from the blocked list and then invoke them to execute malicious behaviors.}
\texttt{cwd} in \texttt{virtual\_cwd\_main\_cwd\_init} holds the path of working directory, which can be used to access other folders by modifying this variable. 
{\bf nginx:} \texttt{cycle-confix.data} temporarily holds the path of configuration file that will be processed by \texttt{ngx\_process\_options}.
{\bf curl:} Variable \texttt{path} holds the location of the data that \texttt{curl} will transfer. By modifying \texttt{path}, attacker can transfer data from an unexpected location, or transfer a malicious file. 
{\bf sqlite3:}  
Variable \texttt{pPath->zOut} holds the location of an opened database; 
variable \texttt{p->sharable} controls if a pBt can be shared with another database. 
Modifying \texttt{iDb} can bypass a lock to a database table.

\section{Related Work}
\label{s:related}

{\bf Defenses on Critical Data. } 
Many solutions add memory safety to unsafe languages  
~\cite{softbound,cets,dfi,ccured,cyclone,ccured}. 
However, enforcing complete memory safety brings significant runtime overhead. 
To achieve a balance between security and overhead, researchers propose selective protection~\cite{xmp,dynpta,cheng2019exploitation}. 
There are some recent works that automatically identify \textit{sensitive} data that have explicit detection rules and could be used in data-only attacks. 
\kenali \cite{kdfi} focuses only on one specific type of critical variable in Kernel code with an explicit and program-specific semantic pattern: ``if a security check fails, it should return a security related error code''. Therefore, it will not be able to generalize to other programs (\eg, user space applications).
DPP \cite{ahmed2023not} uses taint analysis to identify the \textit{sensitive} data and pointer.
\textsc{VIPER}~\cite{ye2023viper} focuses on the syscall-guard variables~\textendash~variables determine to invoke security-related syscalls.
\textsc{STEROIDS}~\cite{pewny2019steroids} leverages the compiler to identify gadgets that can be used in DOP attacks. 
All existing solutions are not able to solve our problem because they are not able to learn high-level semantics and implicit patterns.



\section{Discussion and Conclusion}
\label{sec:con}

In this work, we developed an automated, explicit-pattern-free, 
deep program dependency learning approach to 
significantly reduce the amounts of manual effort in identifying critical variables. 
The evaluation results show that the approach 
can not only re-discover known critical variables,
but also uncover unknown critical variables in the previously investigated programs and other types of programs.
\bibliographystyle{plain}
\bibliography{main}
\end{document}